\documentclass[aps, prb, reprint]{revtex4-1}
\usepackage[utf8]{inputenc}
\usepackage[]{graphicx}

\begin{document}

\title{Validation of a quantized-current source with 0.2~ppm uncertainty}
\date{\today}
\author{Friederike \surname{Stein}}\affiliation{Physikalisch-Technische
Bundesanstalt (PTB), Bundesallee 100, 38116 Braunschweig, Germany.}\noaffiliation
\author{Dietmar \surname{Drung}}\affiliation{Physikalisch-Technische
Bundesanstalt (PTB), Abbestrasse 2-12, 10587 Berlin, Germany.}\noaffiliation
\author{Lukas \surname{Fricke}}\email[electronic address: ]{lukas.fricke@ptb.de}\affiliation{Physikalisch-Technische
Bundesanstalt (PTB), Bundesallee 100, 38116 Braunschweig, Germany.}\noaffiliation
\author{Hansj\"org \surname{Scherer}}\affiliation{Physikalisch-Technische
Bundesanstalt (PTB), Bundesallee 100, 38116 Braunschweig, Germany.}\noaffiliation
\author{Frank \surname{Hohls}}\email[electronic address: ]{frank.hohls@ptb.de}\affiliation{Physikalisch-Technische
Bundesanstalt (PTB), Bundesallee 100, 38116 Braunschweig, Germany.}\noaffiliation
\author{Christoph \surname{Leicht}}\affiliation{Physikalisch-Technische
Bundesanstalt (PTB), Bundesallee 100, 38116 Braunschweig, Germany.}\noaffiliation
\author{Martin \surname{G\"otz}}\affiliation{Physikalisch-Technische
Bundesanstalt (PTB), Bundesallee 100, 38116 Braunschweig, Germany.}\noaffiliation
\author{Christian \surname{Krause}}\affiliation{Physikalisch-Technische
Bundesanstalt (PTB), Bundesallee 100, 38116 Braunschweig, Germany.}\noaffiliation
\author{Ralf \surname{Behr}}\affiliation{Physikalisch-Technische
Bundesanstalt (PTB), Bundesallee 100, 38116 Braunschweig, Germany.}\noaffiliation
\author{Eckart \surname{Pesel}}\affiliation{Physikalisch-Technische
Bundesanstalt (PTB), Bundesallee 100, 38116 Braunschweig, Germany.}\noaffiliation
\author{Klaus \surname{Pierz}}\affiliation{Physikalisch-Technische
Bundesanstalt (PTB), Bundesallee 100, 38116 Braunschweig, Germany.}\noaffiliation
\author{Uwe \surname{Siegner}}\affiliation{Physikalisch-Technische
Bundesanstalt (PTB), Bundesallee 100, 38116 Braunschweig, Germany.}\noaffiliation
\author{Franz J. \surname{Ahlers}}\affiliation{Physikalisch-Technische
Bundesanstalt (PTB), Bundesallee 100, 38116 Braunschweig, Germany.}\noaffiliation
\author{Hans W. \surname{Schumacher}}\affiliation{Physikalisch-Technische
Bundesanstalt (PTB), Bundesallee 100, 38116 Braunschweig, Germany.}\noaffiliation

\begin{abstract}
We report on high-accuracy measurements of quantized current, sourced by a tunable-barrier single-electron pump at frequencies $f$ up to $1$~GHz. The measurements were performed with a new picoammeter instrument, traceable to the Josephson and quantum Hall effects. Current quantization according to $I=ef$ with $e$ the elementary charge was confirmed at $f=545$~MHz with a total relative uncertainty of 0.2 ppm, improving the state of the art by about a factor of 5. For the first time, the accuracy of a possible future quantum current standard based on single-electron transport was experimentally validated to be better than the best realization of the ampere within the present SI.
\end{abstract}
\maketitle

The clocked transfer of single electrons, as implemented in single-electron pumps~\cite{Blumenthal2007,Kaestner2008,Fujiwara2008,Roche2013,Yamahata2014a,Rossi2014} and turnstiles~\cite{Pekola2008}, is regarded as a promising realization of the redefined ampere~\cite{Pekola2013}, which is obtained by fixing the value of the elementary charge $e$~\cite{Mills2011}. Using such a device, the current is realized via $I=-nef$ with $f$ the repetition rate of the charge-transfer cycle and $n$ the number of electrons transferred per cycle. However, in contrast to macroscopic quantum effects like quantum Hall and Josephson effect, the single-electron transfer by periodic manipulation of individual charge carriers can suffer from unwanted tunnelling events. This inevitably results in stochastic deviations from exact quantization~\cite{Kashcheyevs2010,LFricke2013}, leading to a statistical average of transferred electrons per cycle $\left<n\right>$. 

In this paper, we precisely investigate the clocked transfer of electrons using non-adiabatic tunable-barrier pumps~\cite{Blumenthal2007, Kaestner2008} operated at $f\simeq 0.5\ldots1$~GHz, yielding current levels up to $|I|=160$~pA, as required for metrological applications~\cite{Kaestner2012,Giblin2012,Yamahata2014a}. We present direct-current measurements traceable to primary standards using a recently developed ultrastable low-noise current amplifier (ULCA)~\cite{Drung2015,Drung2015a}, excelling the most-accurate current measurement previously conceivable within the present SI~\cite{Jeffery1998,Clothier1989,Funck1991}. 
\begin{figure}[b!]
\includegraphics{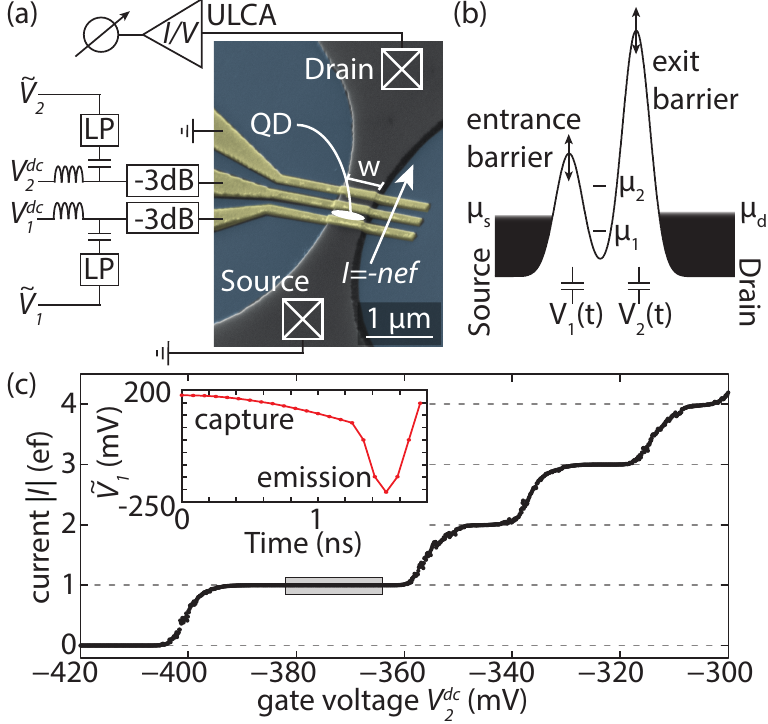}
\caption{(a) False-color SEM micrograph of a single-electron pump together with experimental setup. (b) Sketch of the dynamic quantum dot which is electrostatically defined via topgates. (c) The generated current as a function of control gate voltage takes integer values at multiples of $ef$. As inset, the high-frequency signal applied is shown.}
\label{Fig1}
\end{figure}

The electron pump under investigation (shown in Fig.~\ref{Fig1}(a) as an electron-microscope image) is based on a GaAs/AlGaAs heterostructure comprising a two-dimensional electron gas with carrier density $n_s=2.83\times 10^{15}\ \text{m}^{-2}$ and mobility $\mu=320\ \text{m}^2/\text{Vs}$. A one-dimensional channel is etched with a smoothly tapered constriction of width $w=680$~nm, allowing the operation in perpendicular high magnetic fields as discussed in more detail in e.g. Ref.~\onlinecite{Leicht2011}. A dynamic quantum dot is formed by applying static voltages $V_g^{dc}$ to the Schottky gates $g=1,2$ while the third gate is grounded throughout the measurements. Tunable tunnelling barriers are thereby created between source and drain contacts controlling charge flow as schematically sketched in Fig.~\ref{Fig1}(b). Superposed phase-stable oscillatory signals $\tilde V_g(t)$ generated by a two-channel 12~GS/s arbitrary waveform generator (Tektronix AWG 7122C) are applied via bias tees. The rf signals are low-pass filtered at room temperature using commercial 5.5~GHz low-pass filters (LP, Mini-circuits VLF-5500+) and attenuated by $3$~dB. They drive the clocked transfer of a specific number of electrons $n$ per cycle by tuning both the appropriate tunnelling barrier height as well as the dot's electrochemical potentials $\mu_n$. Electrons are thereby loaded from source onto the dot, isolated from both leads (captured) and finally emitted across the exit barrier to the drain reservoir. 

The measurements are performed in a dilution refrigerator at a base temperature of about $100$~mK and at a constant magnetic field of $B=16$~T applied perpendicular to the sample surface. Current is measured using an ULCA acting as current-to-voltage converter with nominal transresistance of $10^9\ \Omega$ whose output is digitized by an Agilent 3458A voltmeter. We performed measurements on two different samples, denoted A and B. Initially, we will focus on sample A and finally compare the results with data obtained from sample B.

In sample A, the clocked transfer of single electrons is accomplished by  a single-gate drive $\tilde V_1(t)$ ($\tilde V_2(t)\equiv 0$). Thereby, the entrance barrier and the dot's electrochemical potentials are modulated while keeping the height of the exit barrier (defined by the voltage $V_2^{dc}$) fixed. The average number $\left<n\right>$ of transferred electrons per cycle is then predominantly controlled by the voltage $V_2^{dc}$ (defining also the dot's electrochemical potentials at the time of decoupling from source) and shows integer steps corresponding to the transfer of one, two, three or four electrons per cycle on average, as shown in Fig.~\ref{Fig1}(c). Since the error rate in these devices is dominated by errors during the capture phase\cite{Kashcheyevs2010,LFricke2013}, we follow Ref.~\onlinecite{Giblin2012} and slow down the applied waveform during the capture phase as shown in the inset of Fig.~1(c). The capture phase consists of the first fifth of a period of a cosine with frequency $f=150$~MHz, while the emission phase of the waveform, on the other hand, is shortened corresponding to the last three quarters of a cosine of frequency of $f=1.5$~GHz. The nominal repetition rate of the combined pulse pattern is $545.\overline{45}$~MHz.

\begin{figure}[bt]
\includegraphics{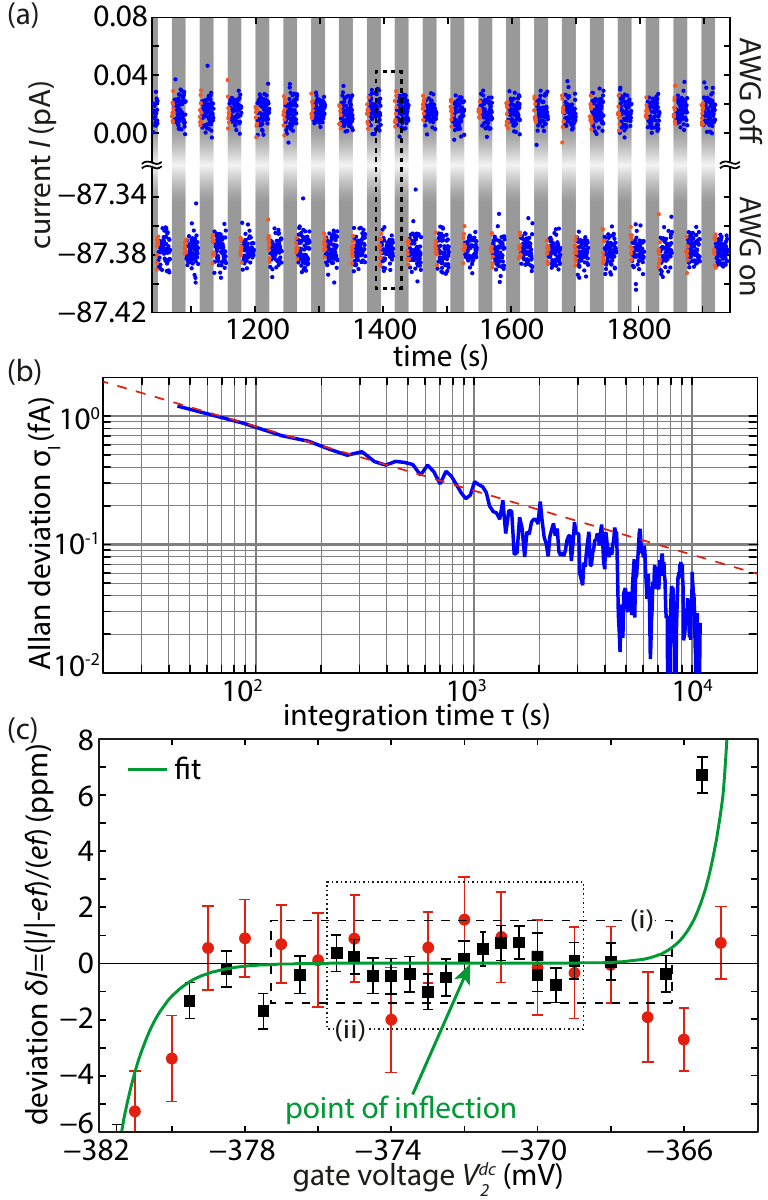}
\caption{Precision measurement of sample A operated at a repetition rate of about $f=545$~MHz (see also Fig.~\ref{Fig1}(c)). (a) Switching measurement of quantized current with gain corrections already included.  (b) Allan deviation of the precision data points derived from the timetrace shown in (a). The red dashed line indicates white noise of $4.5$~fA/$\sqrt{\text{Hz}}$. (c) Gate-voltage dependence of pumped current at $\left<n\right>\approx 1$. Error bars reflect statistical (type A) standard uncertainty ($k=1$). The green solid line represents a fit (see text) from which the point of inflection is derived to $V_2^{dc}\approx -371.9$~mV. Boxes labelled (i),(ii) identify estimates of the plateau extension as discussed in the main text.}
\label{Fig2}
\end{figure}

In order to precisely quantify the charge-transfer process of our dynamic quantum dot, we tune the control voltages in the range of the grey box shown in Fig.~\ref{Fig1}(c). The current measurement is performed on the drain contact with the source lead grounded. To eliminate offsets and drifts caused e.g. by varying thermal voltages, we periodically switch the current on and off by turning on and off the rf signal (43.7~s full-cycle duration). Due to the large electrostatic barriers defining the quantum dot the current is completely blocked in the "off" state and only static offsets remain. A typical result of such a measurement is shown in Fig.~\ref{Fig2}(a) for the dynamic quantum dot tuned to $\left<n\right>\approx 1$. Each data point corresponds to an integration time of $\tau=200$~ms with preceding auto-zero of the voltmeter, resulting in a total acquisition time of about $291$~ms per point (software triggered). Transient effects are omitted by disregarding the first 11 data points (orange points in Fig.~\ref{Fig2}(a), with the first point in each cycle being outside the plot range) after each switching of the signal generator, corresponding to a time interval of about $3.2$~s. Precision data points are then obtained by calculating the difference of the mean of each "on" interval with the means of the neighbouring "off" half intervals as indicated by the dashed box in Fig.~\ref{Fig2}(a).

One of the key features of the ULCA is the temporal stability of its transresistance, which allows measurements over long periods (typical short-term fluctuations over one week are of the order of $0.1$~ppm~\cite{Drung2015}). In Fig.~\ref{Fig2}(b), the Allan deviation as a function of integration time is presented. The red dashed line indicates white noise with $4.5$~fA/$\sqrt{\text{Hz}}$ calculated with eq. (2) in Ref.~\onlinecite{Drung2015}. Taking into account the effective input current noise of the ULCA of $2.4$~fA/$\sqrt{\text{Hz}}$, additional noise from sample and electric wiring is estimated to $3.8$~fA/$\sqrt{\text{Hz}}$. 

\begin{table*}[t!]
\begin{tabular}{|l|p{2.5cm}|l|}
\hline
Contribution (type B) & Rel. uncertainty in ppm ($k=1$) & Derived from\\ \hline
ULCA transresistance & $0.10$ & Calibration against QHR (Interpolation), incl. $T$ correction\\ \hline
Voltmeter gain & $0.08$ & Daily calibration against JVS (Interpolation)\\ \hline
SI value of $e$ & $0.02$ & CODATA 2010~\cite{Mohr2012}\\ \hline
Driving frequency $f$ & $0.01$ & Calibration against $10$~MHz reference\\ \hline
Finite wiring isolation & $0.01$ & Resistance measurements\\ \hline
\bf Total & \bf 0.13 & Root sum of squares \\ \hline
\end{tabular}
\caption{Analysis of the main systematic measurement uncertainty contributions. All represent conservative estimates of upper bounds.}
\label{Tab1}
\end{table*}

Within a period of four months, the transresistance value of the ULCA was calibrated three times traceable to the quantum Hall resistance (QHR) with PTB's 14-bit cryogenic current comparator (CCC)~\cite{Goetz2014,Drung2015a}. Details of the calibration which is performed with a total uncertainty of $0.06$~ppm are described in Refs.~\onlinecite{Drung2015,Drung2015a}. The change in the calibration value observed was $-0.42$~ppm for this specific ULCA during the four-month period. This corresponds to a change of about $-1.3$~ppm per year, slightly better than typical ULCA drift results~\cite{Drung2015}. The temperature inside the amplifier is continuously monitored using an internal temperature sensor for corresponding gain corrections. The voltmeter is gain-calibrated typically once per day against a Josephson voltage standard (JVS) yielding a typical drift of about $0.02$~ppm per day. Table~\ref{Tab1} summarizes the main systematic (type B) measurement uncertainty contributions. The total systematic uncertainty of $0.13$~ppm ($k=1$) is dominated by slow fluctuations and drifts in the ULCA and voltmeter calibration factors.

The detailed investigation of the dependence of clocked current on gate voltage $V_2^{dc}$ is shown in Fig.~\ref{Fig2}(c) as relative deviation from the nominal value, $\delta I=(|I|-ef)/(ef)$. Red data points (circles) correspond to low-accuracy measurements with 75 full cycles of current switching (corresponding to a measurement time of about 55 minutes each) while black data points (squares) represent high-accuracy data with 330 or 500 full switching cycles (equivalent to four or six hours of measurement time, respectively). The data presented in this figure were obtained during a period of one week, demonstrating the stability of the experiment. Error bars represent statistical (type A) uncertainty ($k=1$) only. Depending on the measurement time, combined uncertainties of down to $u_c\approx 0.6$~ppm ($k=1$) per point are obtained. Over a wide range of $V_2^{dc}$ of the order of $10$~mV the data is consistent with the nominal value of $|I|=ef$. 

We model the data using the sum of two single exponential functions (see eq.(3) in Ref.~\onlinecite{Kashcheyevs2014}), assuming that only zero, one, or two electrons are transferred per cycle (which is a valid assumption~\cite{LFricke2013} around $\left<n\right>\approx 1$). From the fit to an extended data set ranging from $(V_2^{dc},\delta I)=(-382\ \text{mV}, -14.3\ \text{ppm})$ to $(-362\ \text{mV}, 835\ \text{ppm})$, we derive the point of inflection at $V_2^{dc}\approx -371.9$~mV which provides the most stable point of operation in terms of variation of gate voltage. Based on this analysis, we develop two criteria to define a region of quantized charge transfer, where the current is independent of the external control parameter within our measurement accuracy. 

Based on the inflection point, criterion (i) defines the current value on the plateau as the mean in an interval of $\pm2$~mV around the inflection point. The plateau extension is then limited by the first data point in each direction which is inconsistent with this value using its expanded type A uncertainty with $k=2$ (dashed box in Fig.~\ref{Fig2}(c)). With this method, we obtain by averaging within this interval $\delta I\pm u_c=(-0.094\pm 0.194)$~ppm with $u_c$ the combined standard uncertainty.

Being completely based on the fit, criterion (ii) defines the extension of the plateau as the interval in which the fit deviates less or equal than $0.01$~ppm from the nominal value (dotted box in Fig.~\ref{Fig2}(c)). In this case, the analysis yields $\delta I\pm u_c=(-0.057\pm 0.205)$~ppm.

As an interim conclusion, the direct-current measurements validate quantized charge transfer using a single-electron pump at $f=545$~MHz (thereby generating a current of about $|I|=87.4$~pA) at an uncertainty of $0.2$~ppm. Regardless of the specific evaluation, the current is consistent with the nominal value within this uncertainty and independent of the external control parameter over a voltage range of about $10$~mV.

We will now present measurements obtained from a second device B. The two samples have different channel widths defined by lithography (device A: $w\approx 680$~nm, device B: $w\approx 570$~nm). Moreover, we use two different schemes of driving the charge-capture process: While sample A was operated rather conventionally using a single-gate signal, sample B is driven by two counter-oscillating sine waves of equal amplitude which are slightly phase-shifted in order to establish uni-directional charge transfer. This leads to a reduction of the plunger action of the entrance gate voltage due to the partial compensation by the modulated exit gate voltage. Moreover, the repetition rate defining the current was increased to $f=1$~GHz. 

Additionally, this measurement shows the capacity of the ULCA setup since now two amplifiers are attached simultaneously to both source and drain contacts. This does not only reduce the type A uncertainty by a factor of $\sqrt 2$, but also allows the independent validation of source and drain currents. Thereby, gate leakage currents flowing from entrance (exit) gate into source (drain) contacts can be excluded within our measurement resolution. Finally, the device and wiring excess noise is reduced compared to sample A to $2.5$~fA/$\sqrt{\text{Hz}}$ leading to an overall noise level of $3.5$~fA/$\sqrt{\text{Hz}}$ for each ULCA channel. Precision data points shown in the following represent the mean of both channels.

\begin{figure}[bt]
\includegraphics{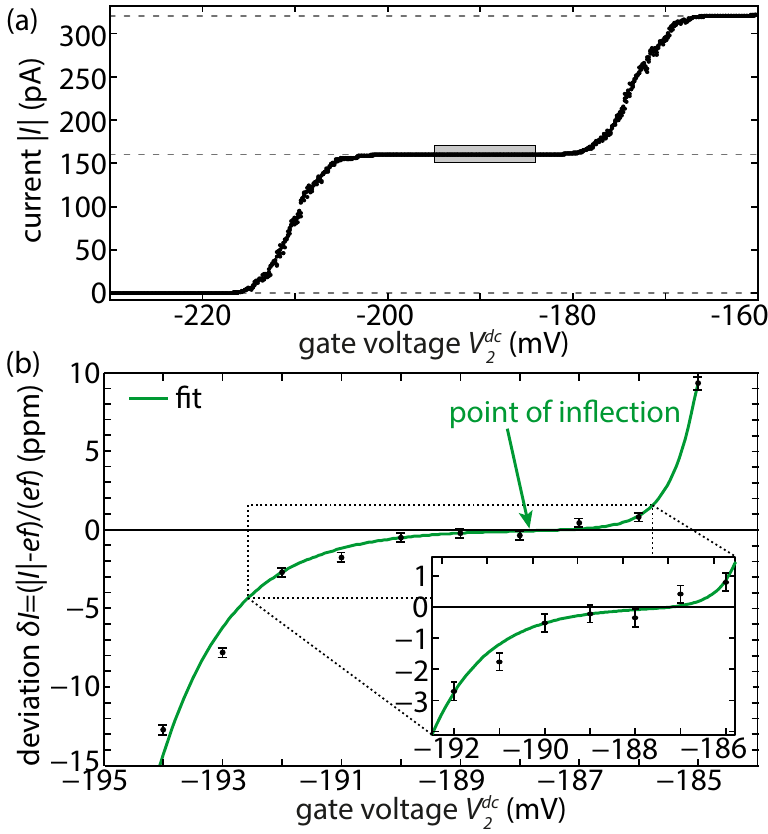}
\caption{Precision measurement of sample B operated at $f=1$~GHz using two phase-shifted sine waves applied to both gates. (a) Low-accuracy direct measurement of quantized current as a function of exit-gate voltage $V_2^{dc}$. (b) Precision data with error bars indicating type A uncertainty ($k=1$), obtained by measuring on both source and drain contacts. The green line indicates a fit similar to Fig.~\ref{Fig2}(c) with the point of inflection at $V_2^{dc}\approx -187.9$~mV. The inset shows a close-up as indicated by the dashed box in the main figure.}
\label{Fig3}
\end{figure}

In Fig.~\ref{Fig3}(a), the parameter dependence of the clocked current as a function of exit gate voltage $V_2^{dc}$ driven at $f=1$~GHz is presented. Plateaus at $|I|\simeq 160$~pA and $|I|\simeq 320$~pA, corresponding to the transfer of $\left<n\right>=1,2$ electrons per cycle, respectively, are well developed. The grey-shaded area marks the range of gate voltages investigated in more detail using the precision measurements displayed in Fig.~\ref{Fig3}(b) following the methods presented for sample A. Each precision data point represents 300 on/off cycles, corresponding to about $3.5$~h of measurement. Error bars reflect type A uncertainty ($k=1$) which is typically $0.29$~ppm per point, yielding a combined uncertainty per data point of about $u_c\approx 0.32$~ppm. The green line is again a fit to a superposition of two single exponential functions, yielding the point of inflection at $V_2^{dc}=-187.9$~mV. However, as evident from the inset, there is still a significant residual slope around this inflection point, thus preventing us from performing the same analysis as for sample A. 
Our results thus demonstrate that fine-structure features of the pump current vs. control parameters, as for instance the residual curvature in the current quantization region, can be investigated with superior accuracy. Note that this residual slope clearly indicates a gate-voltage dependence of $\left<n\right>$ and hence a deviation of current quantization from $ef$ which would not be accessible from lower resolution measurements. Therefore, such measurements support the further development of theoretical models for the underlying transport process as well as the further optimization of pump devices for applications in metrology. 

It is worth noting that the achieved total uncertainty of $0.2$~ppm (obtained for sample A at $f=545$~MHz) is lower than the best ampere realization in the present SI system. Such realization is possible (but, to our knowledge, never was realized) indirectly by combining the SI realizations for resistance and voltage: i) the realization of the Ohm via the Thompson-Lampard calculable capacitor, demonstrated with a total uncertainty of $0.02$~ppm~\cite{Jeffery1998}, and ii) realizations of the volt via a mercury electrometer~\cite{Clothier1989} or via a voltage balance~\cite{Funck1991}, both demonstrated with total uncertainties of $0.27$~ppm. The latter limits the achievable accuracy of this indirect SI ampere realization. 

In summary, the ULCA performance demonstrated here enables a new quality for the analysis of the currents sourced from single-electron pumps. The total uncertainty achieved with our high-accuracy measurements of the direct current sourced by a single-electron pump is about a factor of 5 lower than previously reported uncertainties~\cite{Keller2007,Giblin2012,Bae2015}. It corresponds to an average error rate of $109$~electrons per second. This error rate is well within the achievable bandwidth of today's single-charge detection circuits~\cite{Schoelkopf1998} thus allowing the realization of a high-frequency self-referenced current source~\cite{LFricke2014} with $|I|\sim 100$~pA output current and single-charge error accounting~\cite{Wulf2013}. Such an in-situ validation of high-frequency charge transfer would represent a true quantum standard for the redefined ampere and would also enable a closure of the quantum metrological triangle~\cite{Likharev1985,Scherer2012} by a precision direct-current measurement traceable to JVS and QHR~\cite{Drung2015a}.

The authors acknowledge valuable discussions with V. Kashcheyevs and assistance in clean-room processing by P. Hinze and T. Weimann.
This work was supported by funding from the Joint Research Project "Qu-Ampere" (JRP SIB07) within the European Metrology Research Programme (EMRP). The EMRP is jointly funded by the EMRP participating countries within EURAMET and the European Union.


\begin{thebibliography}{29}%
\makeatletter
\providecommand \@ifxundefined [1]{%
 \@ifx{#1\undefined}
}%
\providecommand \@ifnum [1]{%
 \ifnum #1\expandafter \@firstoftwo
 \else \expandafter \@secondoftwo
 \fi
}%
\providecommand \@ifx [1]{%
 \ifx #1\expandafter \@firstoftwo
 \else \expandafter \@secondoftwo
 \fi
}%
\providecommand \natexlab [1]{#1}%
\providecommand \enquote  [1]{``#1''}%
\providecommand \bibnamefont  [1]{#1}%
\providecommand \bibfnamefont [1]{#1}%
\providecommand \citenamefont [1]{#1}%
\providecommand \href@noop [0]{\@secondoftwo}%
\providecommand \href [0]{\begingroup \@sanitize@url \@href}%
\providecommand \@href[1]{\@@startlink{#1}\@@href}%
\providecommand \@@href[1]{\endgroup#1\@@endlink}%
\providecommand \@sanitize@url [0]{\catcode `\\12\catcode `\$12\catcode
  `\&12\catcode `\#12\catcode `\^12\catcode `\_12\catcode `\%12\relax}%
\providecommand \@@startlink[1]{}%
\providecommand \@@endlink[0]{}%
\providecommand \url  [0]{\begingroup\@sanitize@url \@url }%
\providecommand \@url [1]{\endgroup\@href {#1}{\urlprefix }}%
\providecommand \urlprefix  [0]{URL }%
\providecommand \Eprint [0]{\href }%
\providecommand \doibase [0]{http://dx.doi.org/}%
\providecommand \selectlanguage [0]{\@gobble}%
\providecommand \bibinfo  [0]{\@secondoftwo}%
\providecommand \bibfield  [0]{\@secondoftwo}%
\providecommand \translation [1]{[#1]}%
\providecommand \BibitemOpen [0]{}%
\providecommand \bibitemStop [0]{}%
\providecommand \bibitemNoStop [0]{.\EOS\space}%
\providecommand \EOS [0]{\spacefactor3000\relax}%
\providecommand \BibitemShut  [1]{\csname bibitem#1\endcsname}%
\let\auto@bib@innerbib\@empty
\bibitem [{\citenamefont {Blumenthal}\ \emph {et~al.}(2007)\citenamefont
  {Blumenthal}, \citenamefont {Kaestner}, \citenamefont {Li}, \citenamefont
  {Giblin}, \citenamefont {Janssen}, \citenamefont {Pepper}, \citenamefont
  {Anderson}, \citenamefont {Jones},\ and\ \citenamefont
  {Ritchie}}]{Blumenthal2007}%
  \BibitemOpen
  \bibfield  {author} {\bibinfo {author} {\bibfnamefont {M.~D.}\ \bibnamefont
  {Blumenthal}}, \bibinfo {author} {\bibfnamefont {B.}~\bibnamefont
  {Kaestner}}, \bibinfo {author} {\bibfnamefont {L.}~\bibnamefont {Li}},
  \bibinfo {author} {\bibfnamefont {S.}~\bibnamefont {Giblin}}, \bibinfo
  {author} {\bibfnamefont {T.~J. B.~M.}\ \bibnamefont {Janssen}}, \bibinfo
  {author} {\bibfnamefont {M.}~\bibnamefont {Pepper}}, \bibinfo {author}
  {\bibfnamefont {D.}~\bibnamefont {Anderson}}, \bibinfo {author}
  {\bibfnamefont {G.}~\bibnamefont {Jones}}, \ and\ \bibinfo {author}
  {\bibfnamefont {D.~A.}\ \bibnamefont {Ritchie}},\ }\href {\doibase
  10.1038/nphys582} {\bibfield  {journal} {\bibinfo  {journal} {Nature
  Physics}\ }\textbf {\bibinfo {volume} {3}},\ \bibinfo {pages} {343} (\bibinfo
  {year} {2007})}\BibitemShut {NoStop}%
\bibitem [{\citenamefont {Kaestner}\ \emph {et~al.}(2008)\citenamefont
  {Kaestner}, \citenamefont {Kashcheyevs}, \citenamefont {Amakawa},
  \citenamefont {Blumenthal}, \citenamefont {Li}, \citenamefont {Janssen},
  \citenamefont {Hein}, \citenamefont {Pierz}, \citenamefont {Weimann},
  \citenamefont {Siegner},\ and\ \citenamefont {Schumacher}}]{Kaestner2008}%
  \BibitemOpen
  \bibfield  {author} {\bibinfo {author} {\bibfnamefont {B.}~\bibnamefont
  {Kaestner}}, \bibinfo {author} {\bibfnamefont {V.}~\bibnamefont
  {Kashcheyevs}}, \bibinfo {author} {\bibfnamefont {S.}~\bibnamefont
  {Amakawa}}, \bibinfo {author} {\bibfnamefont {M.~D.}\ \bibnamefont
  {Blumenthal}}, \bibinfo {author} {\bibfnamefont {L.}~\bibnamefont {Li}},
  \bibinfo {author} {\bibfnamefont {T.~J. B.~M.}\ \bibnamefont {Janssen}},
  \bibinfo {author} {\bibfnamefont {G.}~\bibnamefont {Hein}}, \bibinfo {author}
  {\bibfnamefont {K.}~\bibnamefont {Pierz}}, \bibinfo {author} {\bibfnamefont
  {T.}~\bibnamefont {Weimann}}, \bibinfo {author} {\bibfnamefont
  {U.}~\bibnamefont {Siegner}}, \ and\ \bibinfo {author} {\bibfnamefont
  {H.~W.}\ \bibnamefont {Schumacher}},\ }\href {\doibase
  10.1103/PhysRevB.77.153301} {\bibfield  {journal} {\bibinfo  {journal}
  {Physical Review B}\ }\textbf {\bibinfo {volume} {77}},\ \bibinfo {pages}
  {153301} (\bibinfo {year} {2008})}\BibitemShut {NoStop}%
\bibitem [{\citenamefont {Fujiwara}, \citenamefont {Nishiguchi},\ and\
  \citenamefont {Ono}(2008)}]{Fujiwara2008}%
  \BibitemOpen
  \bibfield  {author} {\bibinfo {author} {\bibfnamefont {A.}~\bibnamefont
  {Fujiwara}}, \bibinfo {author} {\bibfnamefont {K.}~\bibnamefont
  {Nishiguchi}}, \ and\ \bibinfo {author} {\bibfnamefont {Y.}~\bibnamefont
  {Ono}},\ }\href {\doibase 10.1063/1.2837544} {\bibfield  {journal} {\bibinfo
  {journal} {Applied Physics Letters}\ }\textbf {\bibinfo {volume} {92}},\
  \bibinfo {eid} {042102} (\bibinfo {year} {2008})}\BibitemShut {NoStop}%
\bibitem [{\citenamefont {Roche}\ \emph {et~al.}(2013)\citenamefont {Roche},
  \citenamefont {Riwar}, \citenamefont {Voisin}, \citenamefont
  {Dupont-Ferrier}, \citenamefont {Wacquez}, \citenamefont {Vinet},
  \citenamefont {Sanquer}, \citenamefont {Splettstoesser},\ and\ \citenamefont
  {Jehl}}]{Roche2013}%
  \BibitemOpen
  \bibfield  {author} {\bibinfo {author} {\bibfnamefont {B.}~\bibnamefont
  {Roche}}, \bibinfo {author} {\bibfnamefont {R.-P.}\ \bibnamefont {Riwar}},
  \bibinfo {author} {\bibfnamefont {B.}~\bibnamefont {Voisin}}, \bibinfo
  {author} {\bibfnamefont {E.}~\bibnamefont {Dupont-Ferrier}}, \bibinfo
  {author} {\bibfnamefont {R.}~\bibnamefont {Wacquez}}, \bibinfo {author}
  {\bibfnamefont {M.}~\bibnamefont {Vinet}}, \bibinfo {author} {\bibfnamefont
  {M.}~\bibnamefont {Sanquer}}, \bibinfo {author} {\bibfnamefont
  {J.}~\bibnamefont {Splettstoesser}}, \ and\ \bibinfo {author} {\bibfnamefont
  {X.}~\bibnamefont {Jehl}},\ }\href {\doibase 10.1038/ncomms2544} {\bibfield
  {journal} {\bibinfo  {journal} {Nature Communications}\ }\textbf {\bibinfo
  {volume} {4}},\ \bibinfo {pages} {1581} (\bibinfo {year} {2013})}\BibitemShut
  {NoStop}%
\bibitem [{\citenamefont {Yamahata}, \citenamefont {Nishiguchi},\ and\
  \citenamefont {Fujiwara}(2014)}]{Yamahata2014a}%
  \BibitemOpen
  \bibfield  {author} {\bibinfo {author} {\bibfnamefont {G.}~\bibnamefont
  {Yamahata}}, \bibinfo {author} {\bibfnamefont {K.}~\bibnamefont
  {Nishiguchi}}, \ and\ \bibinfo {author} {\bibfnamefont {A.}~\bibnamefont
  {Fujiwara}},\ }\href {\doibase 10.1038/ncomms6038} {\bibfield  {journal}
  {\bibinfo  {journal} {Nature Communications}\ }\textbf {\bibinfo {volume}
  {5}},\ \bibinfo {pages} {5038} (\bibinfo {year} {2014})}\BibitemShut
  {NoStop}%
\bibitem [{\citenamefont {Rossi}\ \emph {et~al.}(2014)\citenamefont {Rossi},
  \citenamefont {Tanttu}, \citenamefont {Tan}, \citenamefont {Iisakka},
  \citenamefont {Zhao}, \citenamefont {Chan}, \citenamefont {Tettamanzi},
  \citenamefont {Rogge}, \citenamefont {Dzurak},\ and\ \citenamefont
  {M\"ott\"onen}}]{Rossi2014}%
  \BibitemOpen
  \bibfield  {author} {\bibinfo {author} {\bibfnamefont {A.}~\bibnamefont
  {Rossi}}, \bibinfo {author} {\bibfnamefont {T.}~\bibnamefont {Tanttu}},
  \bibinfo {author} {\bibfnamefont {K.~Y.}\ \bibnamefont {Tan}}, \bibinfo
  {author} {\bibfnamefont {I.}~\bibnamefont {Iisakka}}, \bibinfo {author}
  {\bibfnamefont {R.}~\bibnamefont {Zhao}}, \bibinfo {author} {\bibfnamefont
  {K.~W.}\ \bibnamefont {Chan}}, \bibinfo {author} {\bibfnamefont {G.~C.}\
  \bibnamefont {Tettamanzi}}, \bibinfo {author} {\bibfnamefont
  {S.}~\bibnamefont {Rogge}}, \bibinfo {author} {\bibfnamefont {A.~S.}\
  \bibnamefont {Dzurak}}, \ and\ \bibinfo {author} {\bibfnamefont
  {M.}~\bibnamefont {M\"ott\"onen}},\ }\href {\doibase 10.1021/nl500927q}
  {\bibfield  {journal} {\bibinfo  {journal} {Nano Letters}\ }\textbf {\bibinfo
  {volume} {14}},\ \bibinfo {pages} {3405} (\bibinfo {year}
  {2014})}\BibitemShut {NoStop}%
\bibitem [{\citenamefont {Pekola}\ \emph {et~al.}(2008)\citenamefont {Pekola},
  \citenamefont {Vartiainen}, \citenamefont {M\"ott\"onen}, \citenamefont
  {Saira}, \citenamefont {Meschke},\ and\ \citenamefont {Averin}}]{Pekola2008}%
  \BibitemOpen
  \bibfield  {author} {\bibinfo {author} {\bibfnamefont {J.~P.}\ \bibnamefont
  {Pekola}}, \bibinfo {author} {\bibfnamefont {J.~J.}\ \bibnamefont
  {Vartiainen}}, \bibinfo {author} {\bibfnamefont {M.}~\bibnamefont
  {M\"ott\"onen}}, \bibinfo {author} {\bibfnamefont {O.-P.}\ \bibnamefont
  {Saira}}, \bibinfo {author} {\bibfnamefont {M.}~\bibnamefont {Meschke}}, \
  and\ \bibinfo {author} {\bibfnamefont {D.~V.}\ \bibnamefont {Averin}},\
  }\href {http://dx.doi.org/10.1038/nphys808} {\bibfield  {journal} {\bibinfo
  {journal} {Nature Physics}\ }\textbf {\bibinfo {volume} {4}},\ \bibinfo
  {pages} {120} (\bibinfo {year} {2008})}\BibitemShut {NoStop}%
\bibitem [{\citenamefont {Pekola}\ \emph {et~al.}(2013)\citenamefont {Pekola},
  \citenamefont {Saira}, \citenamefont {Maisi}, \citenamefont {Kemppinen},
  \citenamefont {M\"ott\"onen}, \citenamefont {Pashkin},\ and\ \citenamefont
  {Averin}}]{Pekola2013}%
  \BibitemOpen
  \bibfield  {author} {\bibinfo {author} {\bibfnamefont {J.~P.}\ \bibnamefont
  {Pekola}}, \bibinfo {author} {\bibfnamefont {O.-P.}\ \bibnamefont {Saira}},
  \bibinfo {author} {\bibfnamefont {V.~F.}\ \bibnamefont {Maisi}}, \bibinfo
  {author} {\bibfnamefont {A.}~\bibnamefont {Kemppinen}}, \bibinfo {author}
  {\bibfnamefont {M.}~\bibnamefont {M\"ott\"onen}}, \bibinfo {author}
  {\bibfnamefont {Y.~A.}\ \bibnamefont {Pashkin}}, \ and\ \bibinfo {author}
  {\bibfnamefont {D.~V.}\ \bibnamefont {Averin}},\ }\href {\doibase
  10.1103/revmodphys.85.1421} {\bibfield  {journal} {\bibinfo  {journal}
  {Reviews of Modern Physics}\ }\textbf {\bibinfo {volume} {85}},\ \bibinfo
  {pages} {1421} (\bibinfo {year} {2013})}\BibitemShut {NoStop}%
\bibitem [{\citenamefont {Mills}\ \emph {et~al.}(2011)\citenamefont {Mills},
  \citenamefont {Mohr}, \citenamefont {Quinn}, \citenamefont {Taylor},\ and\
  \citenamefont {Williams}}]{Mills2011}%
  \BibitemOpen
  \bibfield  {author} {\bibinfo {author} {\bibfnamefont {I.~M.}\ \bibnamefont
  {Mills}}, \bibinfo {author} {\bibfnamefont {P.~J.}\ \bibnamefont {Mohr}},
  \bibinfo {author} {\bibfnamefont {T.~J.}\ \bibnamefont {Quinn}}, \bibinfo
  {author} {\bibfnamefont {B.~N.}\ \bibnamefont {Taylor}}, \ and\ \bibinfo
  {author} {\bibfnamefont {E.~R.}\ \bibnamefont {Williams}},\ }\href {\doibase
  10.1098/rsta.2011.0180} {\bibfield  {journal} {\bibinfo  {journal}
  {Philosophical Transactions of the Royal Society A: Mathematical, Physical
  and Engineering Sciences}\ }\textbf {\bibinfo {volume} {369}},\ \bibinfo
  {pages} {3907} (\bibinfo {year} {2011})}\BibitemShut {NoStop}%
\bibitem [{\citenamefont {Kashcheyevs}\ and\ \citenamefont
  {Kaestner}(2010)}]{Kashcheyevs2010}%
  \BibitemOpen
  \bibfield  {author} {\bibinfo {author} {\bibfnamefont {V.}~\bibnamefont
  {Kashcheyevs}}\ and\ \bibinfo {author} {\bibfnamefont {B.}~\bibnamefont
  {Kaestner}},\ }\href {\doibase 10.1103/PhysRevLett.104.186805} {\bibfield
  {journal} {\bibinfo  {journal} {Physical Review Letters}\ }\textbf {\bibinfo
  {volume} {104}},\ \bibinfo {pages} {186805} (\bibinfo {year}
  {2010})}\BibitemShut {NoStop}%
\bibitem [{\citenamefont {Fricke}\ \emph {et~al.}(2013)\citenamefont {Fricke},
  \citenamefont {Wulf}, \citenamefont {Kaestner}, \citenamefont {Kashcheyevs},
  \citenamefont {Timoshenko}, \citenamefont {Nazarov}, \citenamefont {Hohls},
  \citenamefont {Mirovsky}, \citenamefont {Mackrodt}, \citenamefont {Dolata},
  \citenamefont {Weimann}, \citenamefont {Pierz},\ and\ \citenamefont
  {Schumacher}}]{LFricke2013}%
  \BibitemOpen
  \bibfield  {author} {\bibinfo {author} {\bibfnamefont {L.}~\bibnamefont
  {Fricke}}, \bibinfo {author} {\bibfnamefont {M.}~\bibnamefont {Wulf}},
  \bibinfo {author} {\bibfnamefont {B.}~\bibnamefont {Kaestner}}, \bibinfo
  {author} {\bibfnamefont {V.}~\bibnamefont {Kashcheyevs}}, \bibinfo {author}
  {\bibfnamefont {J.}~\bibnamefont {Timoshenko}}, \bibinfo {author}
  {\bibfnamefont {P.}~\bibnamefont {Nazarov}}, \bibinfo {author} {\bibfnamefont
  {F.}~\bibnamefont {Hohls}}, \bibinfo {author} {\bibfnamefont
  {P.}~\bibnamefont {Mirovsky}}, \bibinfo {author} {\bibfnamefont
  {B.}~\bibnamefont {Mackrodt}}, \bibinfo {author} {\bibfnamefont
  {R.}~\bibnamefont {Dolata}}, \bibinfo {author} {\bibfnamefont
  {T.}~\bibnamefont {Weimann}}, \bibinfo {author} {\bibfnamefont
  {K.}~\bibnamefont {Pierz}}, \ and\ \bibinfo {author} {\bibfnamefont {H.~W.}\
  \bibnamefont {Schumacher}},\ }\href {\doibase 10.1103/physrevlett.110.126803}
  {\bibfield  {journal} {\bibinfo  {journal} {Physical Review Letters}\
  }\textbf {\bibinfo {volume} {110}},\ \bibinfo {pages} {126803} (\bibinfo
  {year} {2013})}\BibitemShut {NoStop}%
\bibitem [{\citenamefont {Kaestner}\ \emph {et~al.}(2012)\citenamefont
  {Kaestner}, \citenamefont {Leicht}, \citenamefont {Hohls}, \citenamefont
  {Gotz}, \citenamefont {Drung}, \citenamefont {Pierz}, \citenamefont
  {Ahlers},\ and\ \citenamefont {Schumacher}}]{Kaestner2012}%
  \BibitemOpen
  \bibfield  {author} {\bibinfo {author} {\bibfnamefont {B.}~\bibnamefont
  {Kaestner}}, \bibinfo {author} {\bibfnamefont {C.}~\bibnamefont {Leicht}},
  \bibinfo {author} {\bibfnamefont {F.}~\bibnamefont {Hohls}}, \bibinfo
  {author} {\bibfnamefont {M.}~\bibnamefont {Gotz}}, \bibinfo {author}
  {\bibfnamefont {D.}~\bibnamefont {Drung}}, \bibinfo {author} {\bibfnamefont
  {K.}~\bibnamefont {Pierz}}, \bibinfo {author} {\bibfnamefont {F.~J.}\
  \bibnamefont {Ahlers}}, \ and\ \bibinfo {author} {\bibfnamefont {H.~W.}\
  \bibnamefont {Schumacher}},\ }\href {\doibase 10.1109/cpem.2012.6251125}
  {\bibfield  {journal} {\bibinfo  {journal} {2012 Conference on Precision
  electromagnetic Measurements}\ ,\ \bibinfo {pages} {706}} (\bibinfo {year}
  {2012})}\BibitemShut {NoStop}%
\bibitem [{\citenamefont {Giblin}\ \emph {et~al.}(2012)\citenamefont {Giblin},
  \citenamefont {Kataoka}, \citenamefont {Fletcher}, \citenamefont {See},
  \citenamefont {Janssen}, \citenamefont {Griffths}, \citenamefont {Jones},
  \citenamefont {Farrer},\ and\ \citenamefont {Ritchie}}]{Giblin2012}%
  \BibitemOpen
  \bibfield  {author} {\bibinfo {author} {\bibfnamefont {S.~P.}\ \bibnamefont
  {Giblin}}, \bibinfo {author} {\bibfnamefont {M.}~\bibnamefont {Kataoka}},
  \bibinfo {author} {\bibfnamefont {J.~D.}\ \bibnamefont {Fletcher}}, \bibinfo
  {author} {\bibfnamefont {P.}~\bibnamefont {See}}, \bibinfo {author}
  {\bibfnamefont {T.~J. B.~M.}\ \bibnamefont {Janssen}}, \bibinfo {author}
  {\bibfnamefont {J.~P.}\ \bibnamefont {Griffths}}, \bibinfo {author}
  {\bibfnamefont {G.~A.~C.}\ \bibnamefont {Jones}}, \bibinfo {author}
  {\bibfnamefont {I.}~\bibnamefont {Farrer}}, \ and\ \bibinfo {author}
  {\bibfnamefont {D.~A.}\ \bibnamefont {Ritchie}},\ }\href {\doibase doi:
  10.1038/ncomms1935} {\bibfield  {journal} {\bibinfo  {journal} {Nature
  Communications}\ }\textbf {\bibinfo {volume} {3}},\ \bibinfo {pages} {930}
  (\bibinfo {year} {2012})}\BibitemShut {NoStop}%
\bibitem [{\citenamefont {Drung}\ \emph
  {et~al.}(2015{\natexlab{a}})\citenamefont {Drung}, \citenamefont {Krause},
  \citenamefont {Becker}, \citenamefont {Scherer},\ and\ \citenamefont
  {Ahlers}}]{Drung2015}%
  \BibitemOpen
  \bibfield  {author} {\bibinfo {author} {\bibfnamefont {D.}~\bibnamefont
  {Drung}}, \bibinfo {author} {\bibfnamefont {C.}~\bibnamefont {Krause}},
  \bibinfo {author} {\bibfnamefont {U.}~\bibnamefont {Becker}}, \bibinfo
  {author} {\bibfnamefont {H.}~\bibnamefont {Scherer}}, \ and\ \bibinfo
  {author} {\bibfnamefont {F.~J.}\ \bibnamefont {Ahlers}},\ }\href {\doibase
  10.1063/1.4907358} {\bibfield  {journal} {\bibinfo  {journal} {Review of
  Scientific Instruments}\ }\textbf {\bibinfo {volume} {86}},\ \bibinfo {pages}
  {024703} (\bibinfo {year} {2015}{\natexlab{a}})}\BibitemShut {NoStop}%
\bibitem [{\citenamefont {Drung}\ \emph
  {et~al.}(2015{\natexlab{b}})\citenamefont {Drung}, \citenamefont {G\"otz},
  \citenamefont {Pesel},\ and\ \citenamefont {Scherer}}]{Drung2015a}%
  \BibitemOpen
  \bibfield  {author} {\bibinfo {author} {\bibfnamefont {D.}~\bibnamefont
  {Drung}}, \bibinfo {author} {\bibfnamefont {M.}~\bibnamefont {G\"otz}},
  \bibinfo {author} {\bibfnamefont {E.}~\bibnamefont {Pesel}}, \ and\ \bibinfo
  {author} {\bibfnamefont {H.}~\bibnamefont {Scherer}},\ }\href@noop {}
  {\bibfield  {journal} {\bibinfo  {journal} {IEEE Transactions on
  Instrumentation and Measurement}\ ,\ \bibinfo {pages} {accepted for
  publication}} (\bibinfo {year} {2015}{\natexlab{b}})}\BibitemShut {NoStop}%
\bibitem [{\citenamefont {Jeffery}\ \emph {et~al.}(1998)\citenamefont
  {Jeffery}, \citenamefont {Elmquist}, \citenamefont {Shields}, \citenamefont
  {Lee}, \citenamefont {Cage}, \citenamefont {Shields},\ and\ \citenamefont
  {Dziuba}}]{Jeffery1998}%
  \BibitemOpen
  \bibfield  {author} {\bibinfo {author} {\bibfnamefont {A.}~\bibnamefont
  {Jeffery}}, \bibinfo {author} {\bibfnamefont {R.~E.}\ \bibnamefont
  {Elmquist}}, \bibinfo {author} {\bibfnamefont {J.~Q.}\ \bibnamefont
  {Shields}}, \bibinfo {author} {\bibfnamefont {L.~H.}\ \bibnamefont {Lee}},
  \bibinfo {author} {\bibfnamefont {M.~E.}\ \bibnamefont {Cage}}, \bibinfo
  {author} {\bibfnamefont {S.~H.}\ \bibnamefont {Shields}}, \ and\ \bibinfo
  {author} {\bibfnamefont {R.~F.}\ \bibnamefont {Dziuba}},\ }\href {\doibase
  10.1088/0026-1394/35/2/3} {\bibfield  {journal} {\bibinfo  {journal}
  {Metrologia}\ }\textbf {\bibinfo {volume} {35}},\ \bibinfo {pages} {83}
  (\bibinfo {year} {1998})}\BibitemShut {NoStop}%
\bibitem [{\citenamefont {Clothier}\ \emph {et~al.}(1989)\citenamefont
  {Clothier}, \citenamefont {Sloggett}, \citenamefont {Bairnsfather},
  \citenamefont {Currey},\ and\ \citenamefont {Benjamin}}]{Clothier1989}%
  \BibitemOpen
  \bibfield  {author} {\bibinfo {author} {\bibfnamefont {W.~K.}\ \bibnamefont
  {Clothier}}, \bibinfo {author} {\bibfnamefont {G.~J.}\ \bibnamefont
  {Sloggett}}, \bibinfo {author} {\bibfnamefont {H.}~\bibnamefont
  {Bairnsfather}}, \bibinfo {author} {\bibfnamefont {M.~F.}\ \bibnamefont
  {Currey}}, \ and\ \bibinfo {author} {\bibfnamefont {D.~J.}\ \bibnamefont
  {Benjamin}},\ }\href {\doibase 10.1088/0026-1394/26/1/003} {\bibfield
  {journal} {\bibinfo  {journal} {Metrologia}\ }\textbf {\bibinfo {volume}
  {26}},\ \bibinfo {pages} {9} (\bibinfo {year} {1989})}\BibitemShut {NoStop}%
\bibitem [{\citenamefont {Funck}\ and\ \citenamefont
  {Sienknecht}(1991)}]{Funck1991}%
  \BibitemOpen
  \bibfield  {author} {\bibinfo {author} {\bibfnamefont {T.}~\bibnamefont
  {Funck}}\ and\ \bibinfo {author} {\bibfnamefont {V.}~\bibnamefont
  {Sienknecht}},\ }\href {\doibase 10.1109/tim.1990.1032905} {\bibfield
  {journal} {\bibinfo  {journal} {IEEE Transactions on Instrumentation and
  Measurement}\ }\textbf {\bibinfo {volume} {40}},\ \bibinfo {pages} {158}
  (\bibinfo {year} {1991})}\BibitemShut {NoStop}%
\bibitem [{\citenamefont {Leicht}\ \emph {et~al.}(2011)\citenamefont {Leicht},
  \citenamefont {Mirovsky}, \citenamefont {Kaestner}, \citenamefont {Hohls},
  \citenamefont {Kashcheyevs}, \citenamefont {Kurganova}, \citenamefont
  {Zeitler}, \citenamefont {Weimann}, \citenamefont {Pierz},\ and\
  \citenamefont {Schumacher}}]{Leicht2011}%
  \BibitemOpen
  \bibfield  {author} {\bibinfo {author} {\bibfnamefont {C.}~\bibnamefont
  {Leicht}}, \bibinfo {author} {\bibfnamefont {P.}~\bibnamefont {Mirovsky}},
  \bibinfo {author} {\bibfnamefont {B.}~\bibnamefont {Kaestner}}, \bibinfo
  {author} {\bibfnamefont {F.}~\bibnamefont {Hohls}}, \bibinfo {author}
  {\bibfnamefont {V.}~\bibnamefont {Kashcheyevs}}, \bibinfo {author}
  {\bibfnamefont {E.~V.}\ \bibnamefont {Kurganova}}, \bibinfo {author}
  {\bibfnamefont {U.}~\bibnamefont {Zeitler}}, \bibinfo {author} {\bibfnamefont
  {T.}~\bibnamefont {Weimann}}, \bibinfo {author} {\bibfnamefont
  {K.}~\bibnamefont {Pierz}}, \ and\ \bibinfo {author} {\bibfnamefont {H.~W.}\
  \bibnamefont {Schumacher}},\ }\href {\doibase 10.1088/0268-1242/26/5/055010}
  {\bibfield  {journal} {\bibinfo  {journal} {Semiconductor Science and
  Technology}\ }\textbf {\bibinfo {volume} {26}},\ \bibinfo {pages} {055010}
  (\bibinfo {year} {2011})}\BibitemShut {NoStop}%
\bibitem [{\citenamefont {Mohr}, \citenamefont {Taylor},\ and\ \citenamefont
  {Newell}(2012)}]{Mohr2012}%
  \BibitemOpen
  \bibfield  {author} {\bibinfo {author} {\bibfnamefont {P.~J.}\ \bibnamefont
  {Mohr}}, \bibinfo {author} {\bibfnamefont {B.~N.}\ \bibnamefont {Taylor}}, \
  and\ \bibinfo {author} {\bibfnamefont {D.~B.}\ \bibnamefont {Newell}},\
  }\href {\doibase 10.1103/revmodphys.84.1527} {\bibfield  {journal} {\bibinfo
  {journal} {Review of Modern Physics}\ }\textbf {\bibinfo {volume} {84}},\
  \bibinfo {pages} {1527} (\bibinfo {year} {2012})}\BibitemShut {NoStop}%
\bibitem [{\citenamefont {G\"otz}, \citenamefont {Pesel},\ and\ \citenamefont
  {Drung}(2014)}]{Goetz2014}%
  \BibitemOpen
  \bibfield  {author} {\bibinfo {author} {\bibfnamefont {M.}~\bibnamefont
  {G\"otz}}, \bibinfo {author} {\bibfnamefont {E.}~\bibnamefont {Pesel}}, \
  and\ \bibinfo {author} {\bibfnamefont {D.}~\bibnamefont {Drung}},\ }\href
  {\doibase 10.1109/cpem.2014.6898570} {\bibfield  {journal} {\bibinfo
  {journal} {29th Conference on Precision Electromagnetic Measurements (CPEM
  2014)}\ ,\ \bibinfo {pages} {684}} (\bibinfo {year} {2014})}\BibitemShut
  {NoStop}%
\bibitem [{\citenamefont {Kashcheyevs}\ and\ \citenamefont
  {Timoshenko}(2014)}]{Kashcheyevs2014}%
  \BibitemOpen
  \bibfield  {author} {\bibinfo {author} {\bibfnamefont {V.}~\bibnamefont
  {Kashcheyevs}}\ and\ \bibinfo {author} {\bibfnamefont {J.}~\bibnamefont
  {Timoshenko}},\ }\href {\doibase 10.1109/cpem.2014.6898496} {\bibfield
  {journal} {\bibinfo  {journal} {29th Conference on Precision Electromagnetic
  Measurements (CPEM 2014)}\ ,\ \bibinfo {pages} {536}} (\bibinfo {year}
  {2014})}\BibitemShut {NoStop}%
\bibitem [{\citenamefont {Keller}, \citenamefont {Zimmerman},\ and\
  \citenamefont {Eichenberger}(2007)}]{Keller2007}%
  \BibitemOpen
  \bibfield  {author} {\bibinfo {author} {\bibfnamefont {M.~W.}\ \bibnamefont
  {Keller}}, \bibinfo {author} {\bibfnamefont {N.~M.}\ \bibnamefont
  {Zimmerman}}, \ and\ \bibinfo {author} {\bibfnamefont {A.~L.}\ \bibnamefont
  {Eichenberger}},\ }\href {\doibase 10.1088/0026-1394/44/6/010} {\bibfield
  {journal} {\bibinfo  {journal} {Metrologia}\ }\textbf {\bibinfo {volume}
  {44}},\ \bibinfo {pages} {505} (\bibinfo {year} {2007})}\BibitemShut
  {NoStop}%
\bibitem [{\citenamefont {Bae}\ \emph {et~al.}(2015)\citenamefont {Bae},
  \citenamefont {Ahn}, \citenamefont {Seo}, \citenamefont {Chung},
  \citenamefont {Fletcher}, \citenamefont {Giblin}, \citenamefont {Kataoka},\
  and\ \citenamefont {Kim}}]{Bae2015}%
  \BibitemOpen
  \bibfield  {author} {\bibinfo {author} {\bibfnamefont {M.-H.}\ \bibnamefont
  {Bae}}, \bibinfo {author} {\bibfnamefont {Y.-H.}\ \bibnamefont {Ahn}},
  \bibinfo {author} {\bibfnamefont {M.}~\bibnamefont {Seo}}, \bibinfo {author}
  {\bibfnamefont {Y.}~\bibnamefont {Chung}}, \bibinfo {author} {\bibfnamefont
  {J.~D.}\ \bibnamefont {Fletcher}}, \bibinfo {author} {\bibfnamefont {S.~P.}\
  \bibnamefont {Giblin}}, \bibinfo {author} {\bibfnamefont {M.}~\bibnamefont
  {Kataoka}}, \ and\ \bibinfo {author} {\bibfnamefont {N.}~\bibnamefont
  {Kim}},\ }\href {\doibase 10.1088/0026-1394/52/2/195} {\bibfield  {journal}
  {\bibinfo  {journal} {Metrologia}\ }\textbf {\bibinfo {volume} {52}},\
  \bibinfo {pages} {195} (\bibinfo {year} {2015})}\BibitemShut {NoStop}%
\bibitem [{\citenamefont {Schoelkopf}\ \emph {et~al.}(1998)\citenamefont
  {Schoelkopf}, \citenamefont {Wahlgren}, \citenamefont {Kozhevnikov},
  \citenamefont {Delsing},\ and\ \citenamefont {Prober}}]{Schoelkopf1998}%
  \BibitemOpen
  \bibfield  {author} {\bibinfo {author} {\bibfnamefont {R.~J.}\ \bibnamefont
  {Schoelkopf}}, \bibinfo {author} {\bibfnamefont {P.}~\bibnamefont
  {Wahlgren}}, \bibinfo {author} {\bibfnamefont {A.~A.}\ \bibnamefont
  {Kozhevnikov}}, \bibinfo {author} {\bibfnamefont {P.}~\bibnamefont
  {Delsing}}, \ and\ \bibinfo {author} {\bibfnamefont {D.~E.}\ \bibnamefont
  {Prober}},\ }\href {\doibase 10.1126/science.280.5367.1238} {\bibfield
  {journal} {\bibinfo  {journal} {Science}\ }\textbf {\bibinfo {volume}
  {280}},\ \bibinfo {pages} {1238} (\bibinfo {year} {1998})}\BibitemShut
  {NoStop}%
\bibitem [{\citenamefont {Fricke}\ \emph {et~al.}(2014)\citenamefont {Fricke},
  \citenamefont {Wulf}, \citenamefont {Kaestner}, \citenamefont {Hohls},
  \citenamefont {Mirovsky}, \citenamefont {Mackrodt}, \citenamefont {Dolata},
  \citenamefont {Weimann}, \citenamefont {Pierz}, \citenamefont {Siegner},\
  and\ \citenamefont {Schumacher}}]{LFricke2014}%
  \BibitemOpen
  \bibfield  {author} {\bibinfo {author} {\bibfnamefont {L.}~\bibnamefont
  {Fricke}}, \bibinfo {author} {\bibfnamefont {M.}~\bibnamefont {Wulf}},
  \bibinfo {author} {\bibfnamefont {B.}~\bibnamefont {Kaestner}}, \bibinfo
  {author} {\bibfnamefont {F.}~\bibnamefont {Hohls}}, \bibinfo {author}
  {\bibfnamefont {P.}~\bibnamefont {Mirovsky}}, \bibinfo {author}
  {\bibfnamefont {B.}~\bibnamefont {Mackrodt}}, \bibinfo {author}
  {\bibfnamefont {R.}~\bibnamefont {Dolata}}, \bibinfo {author} {\bibfnamefont
  {T.}~\bibnamefont {Weimann}}, \bibinfo {author} {\bibfnamefont
  {K.}~\bibnamefont {Pierz}}, \bibinfo {author} {\bibfnamefont
  {U.}~\bibnamefont {Siegner}}, \ and\ \bibinfo {author} {\bibfnamefont
  {H.~W.}\ \bibnamefont {Schumacher}},\ }\href {\doibase
  10.1103/PhysRevLett.112.226803} {\bibfield  {journal} {\bibinfo  {journal}
  {Physical Review Letters}\ }\textbf {\bibinfo {volume} {112}},\ \bibinfo
  {pages} {226803} (\bibinfo {year} {2014})}\BibitemShut {NoStop}%
\bibitem [{\citenamefont {Wulf}(2013)}]{Wulf2013}%
  \BibitemOpen
  \bibfield  {author} {\bibinfo {author} {\bibfnamefont {M.}~\bibnamefont
  {Wulf}},\ }\href {\doibase 10.1103/PhysRevB.87.035312} {\bibfield  {journal}
  {\bibinfo  {journal} {Physical Review B}\ }\textbf {\bibinfo {volume} {87}},\
  \bibinfo {pages} {035312} (\bibinfo {year} {2013})}\BibitemShut {NoStop}%
\bibitem [{\citenamefont {Likharev}\ and\ \citenamefont
  {Zorin}(1985)}]{Likharev1985}%
  \BibitemOpen
  \bibfield  {author} {\bibinfo {author} {\bibfnamefont {K.}~\bibnamefont
  {Likharev}}\ and\ \bibinfo {author} {\bibfnamefont {A.}~\bibnamefont
  {Zorin}},\ }\href {\doibase 10.1007/BF00683782} {\bibfield  {journal}
  {\bibinfo  {journal} {Journal of Low Temperature Physics}\ }\textbf {\bibinfo
  {volume} {59}},\ \bibinfo {pages} {347} (\bibinfo {year} {1985})}\BibitemShut
  {NoStop}%
\bibitem [{\citenamefont {Scherer}\ and\ \citenamefont
  {Camarota}(2012)}]{Scherer2012}%
  \BibitemOpen
  \bibfield  {author} {\bibinfo {author} {\bibfnamefont {H.}~\bibnamefont
  {Scherer}}\ and\ \bibinfo {author} {\bibfnamefont {B.}~\bibnamefont
  {Camarota}},\ }\href {\doibase 10.1088/0957-0233/23/12/124010} {\bibfield
  {journal} {\bibinfo  {journal} {Measurement Science and Technology}\ }\textbf
  {\bibinfo {volume} {23}},\ \bibinfo {pages} {124010} (\bibinfo {year}
  {2012})}\BibitemShut {NoStop}%
\end{thebibliography}
\end{document}